\documentstyle[twoside,fleqn,espcrc2,epsf]{article}

\title{Dependence of QCD hadron masses on the number
  of dynamical quarks}

\author{Dong Chen\address{Department of Physics, Columbia University,
	New York, NY 10027, USA}\thanks{The 2 flavor calculation
		in this work was done in collaboration with
		Shailesh Chandrasekharan, Norman H. Christ,
		Weonjong Lee, and Decai Zhu; The 4 and 0 flavor
		calculations were done in collaboration with
		Norman H. Christ and Gregory W.~Kilcup at the Ohio State
		University.  This work was supported in part by the 
		Department of Energy. Presented at Lattice '96.}
	and Robert D.~Mawhinney$^{{\rm a}\,\fnsymbol{footnote}}$}
       
\begin{document}

\def\thepage{CU--TP--776}
\thispagestyle{myheadings}

\begin{abstract}
We have studied the hadron spectrum while varying the number of
light dynamical quarks when the physical lattice spacing and volume are
held fixed relative to the rho mass.  For two and zero flavors of
staggered fermions, we find the nucleon to rho mass ratios
(extrapolated to zero valence quark mass) are very similar.  However,
for four flavors the ratio is 7\% (2 $\sigma$) above the two flavor
result.

\end{abstract}

\maketitle

\section{INTRODUCTION}

The effects of dynamical quarks in zero temperature simulations of QCD
have generally been found to be quite small for ratios of physical
quantities.  In particular, quenched and two flavor dynamical
simulations, for similar lattice spacings and volumes, have yielded
similar results for the nucleon to rho mass ratio.  Currently we can
not probe the differences between zero and two flavor calculations as
the continuum limit is approached, due to limitations in computer
power.  Instead, we have undertaken a comparison of quenched, two
flavor and four flavor QCD at zero temperature, while holding the
lattice spacing and volume fixed in physical units, to look for
dynamical quark effects.

At Lattice `95, the Columbia group reported a spectrum study with two
flavors of staggered quarks on a $16^3\times40$
lattice~\cite{Chen9596}.  Since then, simulations for four and zero
flavors on $16^3\times32$ volumes at Columbia and zero flavors at Ohio
State have been done.  We have chosen our simulation parameters so that
the valence rho mass, extrapolated to zero valence quark mass ($m_{\rm
val} = 0$), is independent of the number of dynamical quarks.  (We
have achieved this to the 3\% level.)   All the simulations consist of
large data samples (by current standards), affording a precise
comparison.

\section{SIMULATION DETAILS}

Table~\ref{tab:parameters} lists the parameters for our simulations.
The Columbia simulations were all done on the 256-node, 16 Gigaflop
computer at Columbia, now completing its seventh year of full-time
calculations.  The two flavor simulation was done with the inexact R
hybrid molecular dynamics algorithm of \cite{Gottlieb87};  the four and
0 flavor simulations used an exact hybrid Monte Carlo algorithm,
employing the $\Phi$ hybrid molecular dynamics algorithm of
\cite{Gottlieb87} and a Monte Carlo accept/reject step.  The quenched
simulations of OSU were done on the Ohio Supercomputer Center T3D using
a mixture of over-relaxed and Metropolis steps.

From Table~\ref{tab:parameters} it is apparent that the two and
four flavor simulations had very similar parameters for the
evolution.  For the four flavor calculation, we used a tighter
conjugate gradient stopping condition in the evolution, so that
the accept/reject step of the exact algorithm would be based
upon as accurate a value for the 5-dimensional Hamiltonian as possible.
The stopping condition we used, and a test condition a factor of
3 smaller, both gave the same sequence of lattices for about
10 time units when evolved from the same starting lattice.

The four flavor run had an acceptance rate of 95\%.  This high
acceptance rate is reassuring for the choice of the running conditions
for our inexact, two flavor run.  In particular this is evidence that
the errors in our two flavor run due to using an inexact algorithm are
small.  This is important, since we want to compare physics from the
two different algorithms.

\begin{table*}[htb]
\caption{Simulation parameters for the four runs presented.  The run
  length, thermalization, hadron measurement frequency and jackknife
  block size are in time units for the CU runs and sweeps for the
  OSU runs.}
\label{tab:parameters}
\begin{center}
\begin{tabular}{|l|c|c|c|c|} \hline
  			& $N_f=4$ (CU) & $N_f=2$ (CU) & $N_f=0$ (CU)
			& $N_f=0$ (OSU)	\\ \hline
  volume		& $16^3\times32$	& $16^3\times40$ 
			& $16^3\times32$	& $16^3\times32$\\\hline
  $\beta$		& 5.4	& 5.7	& 6.05	& 6.05	\\\hline
  $m_{\rm dynamical} a$	& 0.01	& 0.01	& &	\\\hline
 	\hline
  evolution		& HMC	& HMD	& HMC	& OR $+$ Metropolis
    \\\hline
  run length		& 4450	& 4870	& 187,125 & 787,500 \\ \hline
  thermalization	& 250	& 250	& 375	& 25,000 \\ \hline
  acceptance rate 	& 0.95	&	& 0.91	&\\\hline
  trajectory length	& 0.5	& 0.5	& 0.75	&\\\hline
  step size		& 0.0078125	& 0.0078125 & 0.025 &\\\hline
  CG stopping condition	& $1.13\times10^{-6}$	& $1.01\times10^{-5}$
			& &	\\\hline
   total run time	& 5 months	& 7.5 months
			& 1.7 months	& 25k node-hours \\\hline
 	\hline
  hadron source & \multicolumn{4}{c|}{$16^3$ wall, all coordinates
    $(x,y,z)$ even}\\\hline
  valence quark masses & \multicolumn{4}{c|}{0.01, 0.015, 0.02, 0.025}
    \\\hline
  hadron measurement frequency & 5	& 6	& 187.5	& 2500 \\\hline
  number of lattices	& 840	& 770	& 996	& 306\\\hline
  measurements per lattice 	& 4	& 5	& 1 & 4	\\\hline
  jackknife block size	& 50	& 60	& 2250	& 7500\\\hline
  number of blocks	& 84	& 77	& 83	& 102\\\hline
\end{tabular}
\end{center}
\end{table*}

\section{RESULTS}

\begin{figure}[hbt]
\epsfxsize=\hsize
\epsfbox[25 18 587 524]{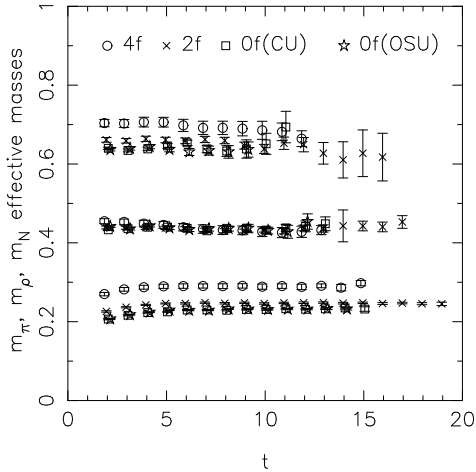}
\caption{The effective masses for the $\pi$, $\rho$ and $N$ for
  $m_{\rm val} = 0.01$.}
\label{fig:effmass}
\end{figure}

Figure \ref{fig:effmass} shows pi, rho and nucleon effective
masses for the four simulations, where the effective masses are found
from fits to time slices $t$ to $t+3$.  The wall source is fixed to
Coulomb gauge and the simple local staggered sinks are used.  While the
rho masses for all cases are very similar, the four flavor nucleon mass
is significantly larger than the others.  Notice there is little
difference between zero and two flavors.

\pagenumbering{arabic}
\addtocounter{page}{1}

\begin{figure}[htb]
\epsfxsize=\hsize
\epsfbox[25 18 587 524]{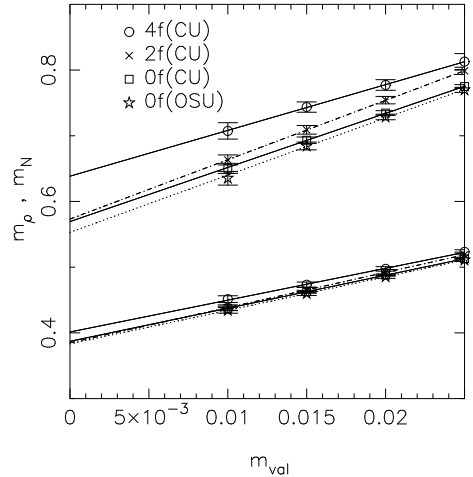}
\caption{$m_{\rho}$ and $m_N$ extrapolated to $m_{\rm val} = 0$.}
\label{fig:n_rho}
\end{figure}

Figure \ref{fig:n_rho} shows the dependence of the rho and nucleon
masses (found from fits between $t=6$ and $t=N_t/2$) on $m_{\rm val}$.
The rho masses agree quite well for all values of $m_{\rm val}$, not
just for $m_{\rm val} \rightarrow 0$.  A consistently larger value for
the nucleon mass for four flavors is apparent.  We have tried breaking
our two flavor data into halves and re-fitting the masses for each
half.  The resulting errors are consistent with the expectation that
they be $\sqrt{2}$ larger.  Similarly for four flavors we compared
results from the first half of our sample with the full sample and the
errors are consistent.  Since we see no evidence for grossly
underestimating our errors for half of our samples, we believe the
errors shown in Figure \ref{fig:n_rho} are reasonably reliable.

\begin{figure}[htb]
\epsfxsize=\hsize
\epsfbox[25 18 587 524]{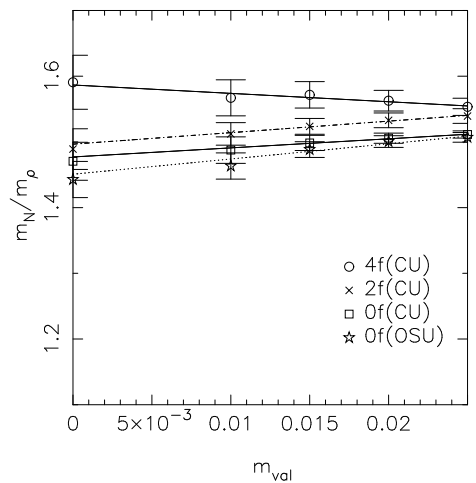}
\caption{$m_N/m_{\rho}$ versus $m_{\rm val}$.  The lines are
 extrapolations in the ratios and the data points at $m_{\rm val}=0$
 are ratios of extrapolated masses.}
\label{fig:n_over_rho}
\end{figure}

Figure \ref{fig:n_over_rho} shows the dependence of the nucleon to rho
mass ratio on $m_{\rm val}$.  The lines are extrapolations of the
ratios, while the data points for $m_{\rm val} = 0$ are the ratio of
the extrapolated masses.  The ratio of the extrapolated masses gives
$m_N/m_{\rho}$ equal to:  1.591(41) for $N_f=4$, 1.489(31) for $N_f=2$,
1.470(29) for CU $N_f=0$ and 1.443(28) for OSU $N_f=0$.  Note the 1
$\sigma$ agreement between the two quenched calculations and the
closeness of the two flavor result.  The four flavor result is 2
$\sigma$ above the two flavor result.

The error on the four flavor ratio of $m_N/m_{\rho}$ is larger than for
two flavors, while the error on the individual masses is quite
similar.  We have studied this in some detail and observe that for four
flavors, for the block size we are using, there is less correlation
between fluctuations in the rho and nucleon propagators than for the
other simulations.  When the ratio is calculated, there is less
cancellation between these uncorrelated fluctuations, leading to the
larger error quoted.

In conclusion, we have seen a 7\% difference (2 $\sigma$) in
$m_N/m_{\rho}$ for $m_{\rm val} = 0$ depending on whether two flavors
or four flavors of dynamical fermions were used in the evolution of the
configurations.  Notice that the increased number of flavors is making
the unphysically large value for the ratio increase.  We are anxious to
understand if this effect persists for larger volumes and when the four
flavor dynamical fermion mass is varied.

We are currently checking for the effects of four dynamical fermions
in other observables.

\end{document}